# LOW SYMMETRY PHASES IN PIEZOELECTRIC SYSTEMS: PZN-xPT SINGLE CRYSTAL AND POWDER

B. NOHEDA, D.E. COX and G. SHIRANE
Physics Dept., Brookhaven National Laboratory, Upton, NY 11973

In this review we address some of the experimental challenges found in the x-ray diffraction characterization of the low-symmetry phases recently discovered in the $Pb(Zr_{1-x}Ti_x)O_3$ (PZT), $Pb(Zn_{1/3}Nb_{2/3})_{1-x}Ti_xO_3$ (PZN-PT) and $Pb(Mg_{1/3}Nb_{2/3})_{1-x}Ti_xO_3$ (PMN-PT) piezoelectric systems around their morphotropic phase boundaries. In particular, we discuss the need for special care in the preparation of powder samples made from single crystals in order to avoid degradation and peak-broadening effects.

## I.  INTRODUCTION

Recently, different low symmetry phases have been observed in both the $Pb(Zr_{1-x}Ti_x)O_3$ (PZT) and the $Pb(Zn_{1/3}Nb_{2/3})_{1-x}Ti_xO_3$ (PZN-PT) ferroelectric systems. A monoclinic phase ($M_A$) was observed in a narrow region of compositions of the PZT phase diagram, between the rhombohedral (R) and tetragonal (T) phases, at the morphotropic phase boundary (MPB) [1-3]. Furthermore, it was also reported that this region is irreversibly enlarged towards the rhombohedral side by poling [4]. The enhancement of the piezoelectric response observed in the vicinity of the $M_A$ phase can now be understood by considering that the polarization vector in this phase can rotate within the monoclinic plane [5]. Studies of single crystals of the high-strain piezoelectric system

PZN-PT have also revealed the existence of a lower-symmetry orthorhombic phase (O) between the well-established R and T phases [6-8]. When an electric field is applied to rhombohedral PZN-4.5%PT along the [001] direction, a tetragonal phase can be induced in the crystal at sufficiently high fields [9]. At intermediate field values the crystal has monoclinic $M_A$ symmetry [10], as in the PZT case, as the polarization rotates from [111] to [001], in agreement with first-principles calculations [11]. However, for PZN-8%PT, which lies closer to the MPB, a different monoclinic phase ($M_C$) can be induced, in which the polarization rotates between the orthorhombic [101] and tetragonal [001] polar axes [12]. The $M_A$ and $M_C$ monoclinic phases are two of the three possible monoclinic phases deduced for ferroelectric perovskites from an eighth-order Devonshire approach [13]. First-principles calculations also predict the existence of a R-$M_A$-$M_C$-T path in the case of rhombohedral PZT under a [001] electric field [14], similar to what was observed for PZN-8%PT [12,15].

During the diffraction experiments mentioned above we have found experimental challenges that are linked to the nature of these piezoelectric systems, in particular, to the near-degeneracy of the different phases close to the MPB and the important role played by the internal strain fields and the mechanical boundary conditions. Thus, in this review we will not show the important results that are already published. Instead, we want to emphasize the importance of the sample preparation, which we have used as a stepping-stone to help us obtain a clearer picture of the long-range structure in these materials.

Single crystals of PZN-4.5PT, PZN-8%PT and PZN-9%PT, as described in refs. 12, 10 and 6, respectively, were used in the present work. The crystals were cut into cubes with at least two of their faces perpendicular to a cubic <100> direction. The crystals were poled either with a field of about 20kV/cm at room temperature, or field-cooled from high temperature under a smaller field of about 10kV/cm.

The powder diffraction measurements were made at the Brookhaven National Synchrotron Light Source at beamline X7A at a wavelength of about 0.7Å, using either a position-sensitive-detector (PSD) or a crystal-analyzer, as described in refs. [1-3,6,8]. The single crystal experiment described in this work was also performed at the same beamline. For the powder diffraction experiments, a small piece of the crystal was lightly crushed in an agate mortar under acetone. The fraction of material retained between sieves with mesh sizes of 38 and 44 μm was loaded into a 0.2 mm glass capillary, as described in ref. [6].

We use the term "PASC" ("properly-arranged single crystals") for such samples, which differ from normal powder samples in that the number of grains in the beam is relatively small and good powder-averaging is not always possible, especially if the capillary is inside a cryostat and cannot be rotated. Because there is minimal degradation of the peak widths, this technique is excellent for characterizing the unit cell symmetry but may not be suitable for full structural analysis. For the latter purpose both PASC and normal powder samples should be used.

### III. PZN-4.5%PT

Figure1 shows part of the diffraction pattern collected with the PSD at room temperature from a PASC powder sample prepared from a crystal previously poled under a field of 15 kV/cm along the [001] direction. From the peak profiles of the pseudo-cubic (200) and (222) reflections shown in the insets, the symmetry is clearly seen to be rhombohedral.

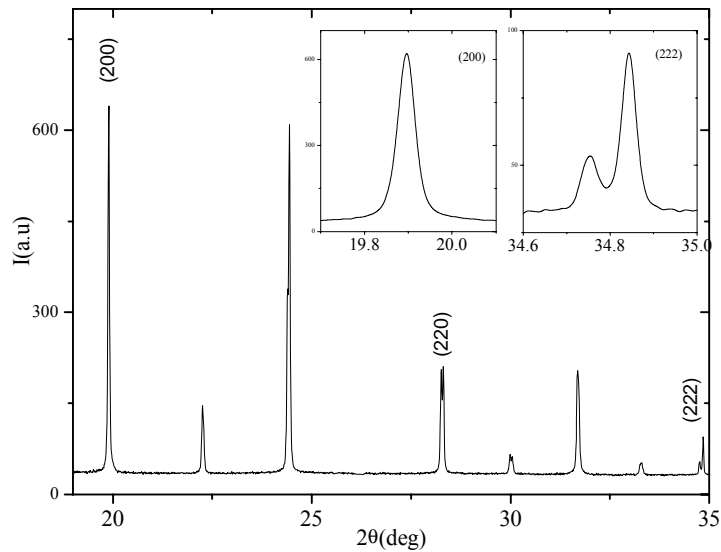

FIGURE 1. Diffraction pattern from a poled PZN-4.5%PT capillary sample obtained with the PSD at $\lambda = 0.70071$Å. The rhombohedral lattice parameters are $a = 4.0567(1)$ Å and $\alpha = 89.893(1)$ deg.

The evolution of the pseudo-cubic (220) reflection as a function of temperature obtained with crystal-analyzer geometry is shown in Figure 2. On heating, the splitting between the two peaks (which reflects the rhombohedral distortion) decreases until a single sharp peak is present at approximately 165°C, which is ≈ 1°C above the rhombohedral-cubic transition temperature [16]. On cooling, the peak broadens gradually, but he splitting is not recovered at room temperature, indicative of a much smaller domain size and recovery of the relaxor state of the as-grown unpoled crystal.

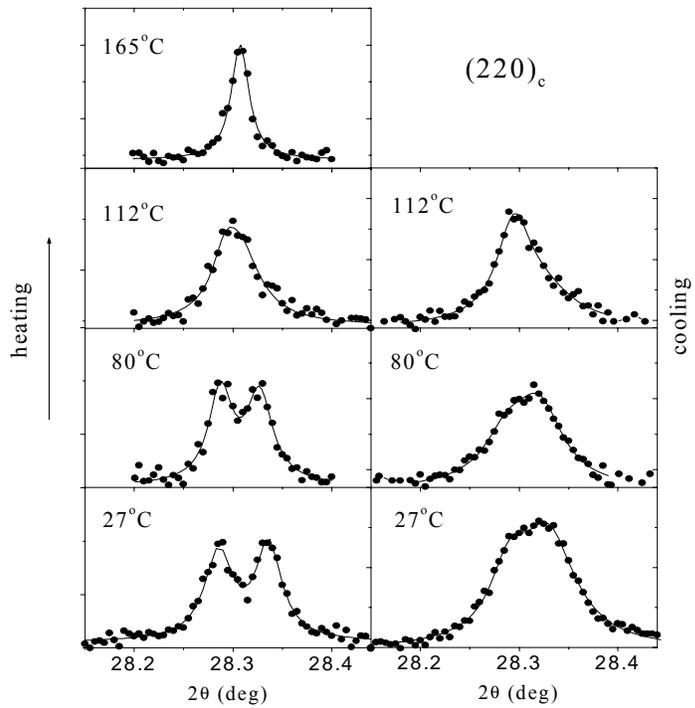

FIGURE 2. Temperature evolution of the $(220)_{pc}$ reflection in PZN-4.5%PT.

## IV. PZN-8%PT

Data were collected with crystal-analyzer geometry from a PASC powder sample prepared from an as-grown (unpoled) PZN-8%PT single crystal. At 250°C, the first four reflections were all sharp single peaks (FWHM's ≈ 0.01-0.02°), characteristic of cubic symmetry. The evolution of the (200) peak profile was followed as a function of decreasing temperature. This peak remained single down to 170°C but between 160°C-110°C, additional peaks were observed on either side of the cubic peak, consistent with the appearance of a tetragonal phase. Between 160°C-110°C the amount of tetragonal phase grew from about 30% to 70%, the c/a ratio increased from 1.0041 to 1.0071, while the cubic lattice parameter remained essentially constant as seen in Figure3. Three peaks were also observed between 110°C-30°C, but there is a distinct jump in the "cubic" lattice parameters, which could be interpreted as a cubic-rhombohedral transition. However, there is also a steady decrease in the "tetragonal" $c$ parameter, which would be most unusual in a tetragonal ferroelectric system, and we believe a more plausible explanation is that a transformation to monoclinic symmetry has occurred at about 90°C. The decrease in $c$ would then be consistent with a rotation of the polar axis away from the [001] direction.

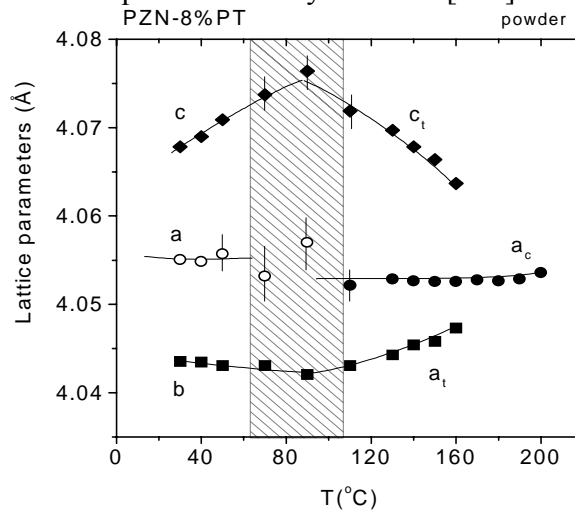

FIGURE 3. Temperature evolution of the lattice parameters derived from the pseudo-cubic (200) reflection from a PZN-8%PT capillary sample prepared from an as-grown crystal.

Data were also collected from a 2x2x2mm$^3$ single crystal which had previously been poled along [001]. The crystal had gold electrodes with wires attached so that an electric field could be applied *in-situ* [12]. The peak profiles obtained from the (h00) reflections perpendicular to the field direction are shown in Figure4. These profiles are characterized by a very broad (h00) peak on the low-angle side and a

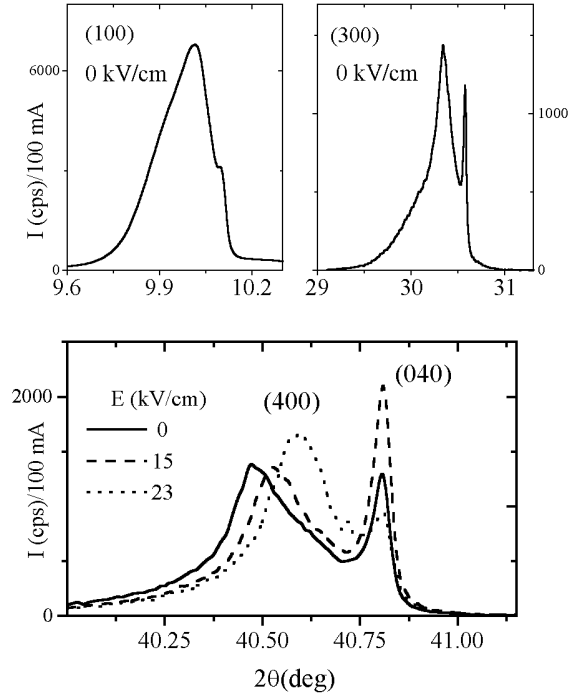

FIGURE 4 (h00) peaks from a PZN-8%PT single crystal poled along [001] with scattering vector perpendicular to the field direction. At the bottom, the evolution of the (400) and (040) reflections with electric field is shown. These data were obtained during the measurements reported in ref.[12].

sharp (0h0) peak on the high-angle side, suggesting the existence of two phases with very different coherence lengths [17]. However, subsequent experiments with high-energy x-rays showed that the broadening of the low-angle peak is a "skin effect" due to inhomogeneities near to the surfaces [15], and that the two peaks actually correspond to the monoclinic parameters $a_m$ and $b_m$ [12]. At the bottom of Figure 4, the evolution of (400) and (040) with electric field is shown, demonstrating that while $b_m$ remains constant with increasing field, $a_m$ decreases, eventually becoming equal to $b_m$ when the tetragonal phase is reached.

Figure 5 shows powder data collected at room temperature from a PASC sample prepared from a PZN-8%PT single crystal previously poled along the [001] direction. From the splitting of the pseudo-cubic

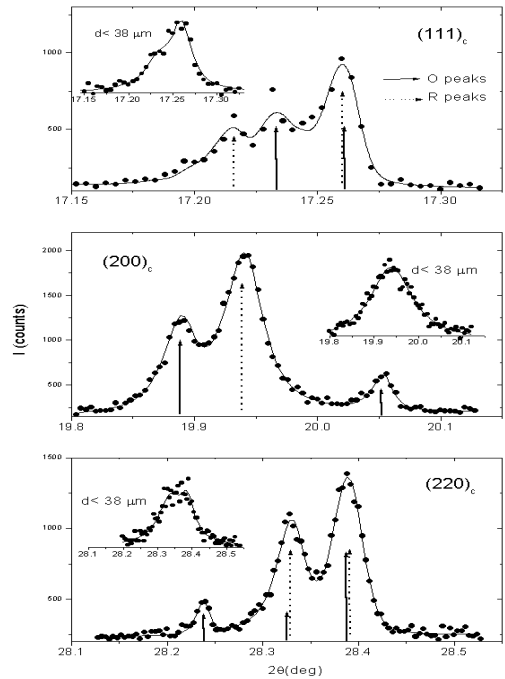

FIGURE 5. (111), (200) and (220) pseudo-cubic reflections of a powder made from a PZN-%8PT crystal poled along [001]. The larger plots show the data from the PASC sample (38-44 μm); the insets show the same reflections taken on a normal sample with smaller grains.

(111), (200) and (220) reflections, it is straightforward to deduce that the sample consists of two components; an orthorhombic phase with $a$ = 5.7513 Å, $b$ = 4.0301 Å, and $c$ = 5.7364 Å, and a rhombohedral phase with $a$ = 4.0567 Å and $\alpha$ = 89.89°, in the approximate ratio 1:1. This orthorhombic phase is similar to that recently reported by Cox et al. [6] which is B-centered. It is interesting to note that this B-centered cell is equivalent to a primitive monoclinic cell with $a = c$ = 4.061 Å, $b$ = 4.030 Å, $\beta$ = 90.15°. Single crystal experiments have shown that after poling PZN-8%PT is purely orthorhombic [12] so the second rhombohedral phase, similar to that of the unpoled crystal, appears to have been recovered by grinding. Also shown for comparison in the insets to Figure 5 are data collected from a normal powder sample with smaller grains made from the same single crystal. The profiles in this case exhibit broad envelopes in which the individual peaks are not resolved.

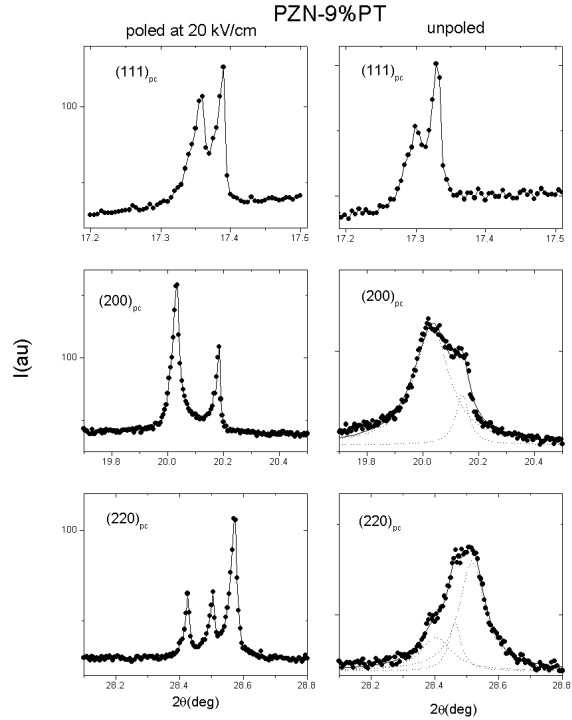

FIGURE 6. (111), (200) and (220) pseudo-cubic reflections of poled (left side) and unpoled (right side) PZN-9%PT PASC samples, made from a PZN-9%PT single crystal [6].

## V. PZN-9%PT

Figure 6 shows data collected from two different PASC samples from the same PZN-9%PT crustal poled along [001]. This composition shows pure orthorhombic symmetry as described in refs. [5,6]. Figure 6 (left side) shows the powder data from the part of the sample below the electrode, the poled sample. The sharpness of the peaks and the clearly orthorhombic splitting show the high quality of sample, which was not at all degraded after crushing and sieving. Figure 6 (right side) shows the data from the same sample, but this time from a piece of the crystal that was not poled, outside the electrode area. It can be seen that the peaks are much broader, corresponding to a more disordered sample. However, from a comparison of the two sets of data in Figure 6, it is clear that the same underlying symmetry is present in both patterns. Thus, in this region of composition, poling does not change the symmetry of the crystal (considerably higher fields are needed to induce the tetragonal phase), but simply orders it. Very recently, the "PASC" technique has also been used by Ye et al.[18] to study PMN-35%PT.

Acknowledgments
We wish to thank all our collaborators, in particular, L.E. Cross, M. Durbin, S-E. Park, P. Rehrig, Y. Uesu, T. Vogt and Z-G. Ye for very helpful discussions. Financial support by the U.S. Department of Energy, (contract No. DE-AC02-98CH10886) is also acknowledged.

References
1. B. Noheda et al., App. Phys. Lett. **74**, 2059 (1999)
2. B. Noheda et al., Phys. Rev. B **61**, 8686 (1999)
3. B. Noheda et al., Phys. Rev. B **63**, 14103 (2001)
4. R. Guo et al., Phys. Rev. Lett. **84**, 5216 (2000)
5. L. Bellaiche et al. Phys. Rev. Lett. **84**, 5427 (2000)
6. D.E. Cox. et al. Appl. Phys. Lett. **79**, 400 (2001)
7. Y. Uesu et al. J. Phys. Soc. Japan (submitted). <cond-mat/0106552>
8. D. La-Ourattapong et al., (submitted). <cond-mat/0108264>
9. S.E. Park and T. Shrout J. Appl. Phys.**82**, 1804 (1997)
10. B. Noheda et al. (in preparation)
11. H. Fu and R. Cohen Nature **403**, 281 (2000)
12. B. Noheda et al., Phys. Rev. Lett. **86**, 3891 (2001)
13. D. Vanderbilt and M. Cohen, Phys. Rev. B. **63**, 094108 (2001)
14. L. Bellaiche et al., Phys. Rev. B. **64**, 060103(R) (2001)
15. K. Ohwada et al., J. Phys.Soc.Japan (in press) <cond-mat/0105086>
16. J. Kuwata et al., Jpn. J. Appl. Phys. **21**, 1298 (1982).
17. M. K. Durbin et al., J. Appl. Phys. **87**, 8159 (2000).
18. Z. Ye et al. , Phys. Rev. B **64**, 1841XX (2001) (in press)